\documentclass[twocolumn,tighten]{aastex631}

\usepackage{amsmath}

\bibpunct{(}{)}{;}{a}{}{,}

\newcommand{\ee}[1]{\mbox{${} \times 10^{#1}$}}

\newcommand{\kms}{\mbox{km s$^{-1}$}}
\newcommand\cmv{\mbox{cm$^{-3}$}}
\newcommand\cmc{\mbox{cm$^{-2}$}}

\newcommand{\x}{\mbox{${}\times{}$}}

\newcommand{\sfrth }{\mbox{${\rm SFR}_{\rm th}$}}
\newcommand{\sfrff }{\mbox{${\rm SFR}_{\rm th, ff}
$}}
\newcommand{\sfrobs }{\mbox{${\rm SFR}_{\rm obs}$}}

\newcommand{\msun}{\mbox{M$_\odot$}}

\newcommand{\n}{\mbox{$n$}}

\newcommand{\mco}{\mbox{$M_{\rm CO}$}} 
\newcommand{\lco}{\mbox{$L_{\rm CO}$}} 
\newcommand{\ico}{\mbox{$W_{\rm CO}$}} 
\newcommand{\xco}{\mbox{$X_{\rm CO}$}} 
\newcommand{\xcoz}{\mbox{$X_{\rm CO, 0}$}} 
\newcommand{\xcounit}{\mbox{$\cmc ({\rm K}\ \kms)^{-1}$}} 
\newcommand{\epsff}{\mbox{$\epsilon_{\rm ff}$}} 
\newcommand{\epsffobs}{\mbox{$\epsilon_{\rm ff, obs}$}} 
\newcommand{\epscs}{\mbox{$\epsilon_{\rm cs}$}} 
\newcommand{\thdobs}{\mbox{$Q_{\rm th}$}} 
\newcommand{\thdobsff}{\mbox{$Q_{\rm ff}$}} 
\newcommand{\thdobsstep}{\mbox{$Q_{\rm step}$}} 
\newcommand{\tff}{\mbox{$t_{\rm ff}$}} 
\newcommand{\alphavir}{\mbox{$\alpha_{\rm vir}$}} 
\newcommand{\alphavirz}{\mbox{$\alpha_{\rm vir, 0}$}} 
\newcommand{\alphaco}{\mbox{$\alpha_{\rm CO}$}} 
\newcommand{\alphacoz}{\mbox{$\alpha_{\rm CO, 0}$}} 
\newcommand{\muhh}{\mbox{$\mu_{\rm H_{2}}$}} 
\newcommand{\nhh}{\mbox{$N_{\rm H_{2}}$}} 
\newcommand{\hii}{\mbox{\ion{H}{2}}} 
\newcommand{\rbeam}{\mbox{$r_{\rm beam}$}} 

\newcommand{\mean}[1]{\mbox{$\langle#1\rangle$}} 

\newcommand{\hh}{\mbox{{\rm H}$_2$}}

\newcommand{\coo}{\mbox{$^{13}$CO}}

\newcommand{\jj}[2]{\mbox{$J = #1\rightarrow#2$}}

\newcommand{\sigmam}{\mbox{$\Sigma_{\rm mol}$}}
\newcommand{\go}{\mbox{$G_0$}}

\newcommand{\kkms}{\mbox{K\ \kms}}
\newcommand{\alphacounit}{\mbox{\msun (K\ \kms\ pc$^2$)$^{-1}$}}

\newcommand{\rgal}{\mbox{$R_{\rm gal}$}}
\newcommand{\rgalsun}{\mbox{$R_{\rm gal, \odot}$}}

\newcommand{\msunyr}{\mbox{M$_\odot$ yr$^{-1}$}}
\newcommand{\msunpc}{\mbox{M$_\odot$ pc$^{-2}$}}

\newcommand{\pc}{\,{\rm pc}}

\shorttitle {Theory Meets Observations}
\shortauthors{Evans, Kim, and Ostriker}

\begin{document}

\setcounter{table}{0}

\title{Slow Star Formation in the Milky Way: Theory Meets Observations}

\correspondingauthor{Neal J. Evans II}
\email{nje@astro.as.utexas.edu}

\author[0000-0001-5175-1777]{Neal J. Evans II}
\affiliation{Department of Astronomy, The University of Texas at Austin,
2515 Speedway, Stop C1400, Austin, Texas 78712-1205, USA}

\author[0000-0001-6228-8634]{Jeong-Gyu Kim}
\affiliation{Korea Astronomy and Space Science Institute, Daejeon 34055, Republic Of Korea}
\affiliation{Department of Astrophysical Sciences, Princeton University, Princeton, NJ, 08544, USA}

\author[0000-0002-0509-9113]{Eve C.~Ostriker}
\affiliation{Department of Astrophysical Sciences, Princeton University, 
Princeton, NJ, 08544, USA}

\begin{abstract}
The observed star formation rate of the Milky Way can be explained by applying
a metallicity-dependent factor to convert CO luminosity to molecular gas mass and a star formation efficiency per free-fall time that depends on the virial parameter of a molecular cloud. 
These procedures also 
predict the trend of star formation rate surface density with Galactocentric radius. 
The efficiency per free-fall time variation with virial parameter
plays the major role in bringing theory into agreement with observations for
the total star formation rate, while the metallicity dependence of the
CO luminosity to mass conversion is most notable in the variation with 
Galactocentric radius.
Application of these changes resolves a factor of over 100 discrepancy 
between observed and theoretical star formation rates
that has been known for nearly 50 years.
\end{abstract}


\section{Introduction}\label{s:intro}

The observed star formation rate in the Milky Way (\sfrobs), 
averaged over recent history of the Galaxy, is
estimated to be $1.65$--$1.9$ \msunyr\ \citep{2015ApJ...806...96L, Chomiuk11}. 
In contrast, the SFR
predicted if all the clouds identified in CO surveys are collapsing at
free-fall exceeds the observed rate by at least two orders of magnitude.
With a total molecular mass of 1\ee9 \msun\
\citep{2015ARA&A..53..583H}
and a free-fall time of 3.34\ee6 yr, 
taking a characteristic density of $100$ \cmv, 
if all molecular gas ($M_{\rm mol,tot}$) forms stars 
with complete efficiency in a free-fall time ($t_{\rm ff,mol}$), 
the free-fall star formation rate, 
$\sfrff \equiv M_{\rm mol,tot}/t_{\rm ff,mol} = 300$ \msunyr\
\citep{2021ApJ...920..126E}. We
characterize this issue by $\thdobs = \sfrth/\sfrobs$, 
the ratio of the SFR predicted
by a given theory to the observed SFR. Thus, for the Galaxy, 
$\thdobsff = 158-182$.

This problem can be restated as the slowness of star formation
\citep[see e.g., reviews of][]{2007ARA&A..45..565M,2014prpl.conf..243K, 2014prpl.conf...77P}, with a need to explain why the efficiency per free-fall time
$\epsffobs\equiv\sfrobs/\sfrff$  is at most only a few percent. As noted above, the average $\epsffobs=1/\thdobsff$ needed to
bring current estimates of molecular cloud properties and star formation rates
into agreement on a Galaxy-scale level is even lower:
$\epsffobs = 0.006$.
Alternatively, the molecular gas depletion time is $t_{\rm dep} \equiv M_{\rm mol,tot}/{\rm SFR_{obs}} = 0.5$--$0.6$ Gyr, longer
than the free-fall time by the factor $\thdobsff$.

This huge discrepancy captured by \thdobsff\  is one of the oldest \citep{1974ARA&A..12..279Z,
  1974ApJ...192L.149Z} and most embarrassing in the field of star formation.
It has been identified as the first of the three ``big problems" in star formation,
along with understanding stellar clustering and the origin of the initial mass function 
\citep{2014PhR...539...49K}.

The problem cannot be solved by rotational stabilization as
rotational energies are far less than gravitational or turbulent energies
(\citealt{2020A&A...633A..17B} and references therein). Some combination
of magnetic fields, turbulence, and feedback is generally invoked to explain why star formation is slow, but simulations 
with comparable gravitational and turbulent energy 
have difficulty matching the observations 
($\epsffobs = 0.006$),
instead producing $\epsff\gtrsim 0.1$, 
unless turbulence is continuously driven (with an artificial stirring force) and/or very strong magnetic fields are included
\citep[e.g.][]{2012ApJ...759L..27P,Federrath15b,2016ApJ...829..130R,2018MNRAS.476..771C,2021ApJ...911..128K}.
However, both theory and simulations suggest that $\epsff$ drops steeply with increasing virial parameter \citep[e.g.][]{2005ApJ...630..250K,2008MNRAS.386....3C,2011ApJ...730...40P,2012ApJ...759L..27P,2012ApJ...761..156F,2021ApJ...911..128K}, and recent observations suggest that the relative importance of gravity compared to turbulence in GMCs is less than traditionally thought \citep[e.g.,][]{2020ApJ...901L...8S,2021ApJ...920..126E}.

Reproducing the star formation rate of the whole Galaxy provides
the most stringent test of theory. Comparisons to samples of clouds leave
open issues of sample selection and time varying efficiency
\citep{2019ARA&A..57..227K}. 
Because observations
sample clouds at an undetermined time in their history, large variations
in star formation efficiency are observed from cloud to cloud
\citep[e.g.,][]{2016ApJ...831...73V,2016ApJ...833..229L}.
Simulations show that due to the expansion of clouds produced by feedback, the instantaneous measured $\epsff$ is positively  correlated with the instantaneous virial parameter, whereas $\epsff$ is inversely correlated with the virial parameter prior to the onset of star formation  \citep[][Fig. 16 vs. Fig 15]{2021ApJ...911..128K}.
Simulations also show that various star formation tracers systematically
over-predict and under-predict the actual \epsff\ at different times in
the cloud history
\citep[e.g., Fig. 7 in][]{2022arXiv220100882G}. 
In contrast, the whole-Galaxy star formation rate based on young populations (like \hii\ regions) averages over all clouds over the last 5-10 Myr
and is quite well determined. Any credible star formation theory must
be able to predict the observed value within reasonable uncertainties.

Furthermore, star formation in the Milky Way is not uniquely slow, lying
near the Kennicutt-Schmidt relation between star formation rate and gas
surface densities
\citep[Fig. 11 in][]{2012ARA&A..50..531K}. 
In a study of
14 nearby galaxies analyzed with the ``Milky Way" conversion from 
CO luminosity to molecular mass, including all the CO emission,
\citet{2018ApJ...861L..18U} 
found a median over all lines of sight of $\epsffobs = 0.007 \pm 0.003$,
essentially identical to the Milky Way. The solution of the problem for the
Milky Way may point the way to better understanding of star-forming
galaxies more generally. 

In this paper, we take a fresh look at this problem by reconsidering
the determinations of masses of molecular clouds, and using measurements
of \epsff\ from MHD simulations with different virial parameters in combination with estimates of observed virial parameters in Milky Way clouds.

\section{Observational Constraints}\label{obs}

The basic constraints are the observed star formation rate and the observed
properties of molecular structures (CO luminosity, size, velocity dispersion)
that allow us to compute mass, free-fall time, and virial parameter.

For our purposes, the relevant estimates of SFR should average over times
similar to lifetimes of molecular clouds, 
several Myr \citep{2015ARA&A..53..583H}.
\citet{Chomiuk11} collected such estimates of the total SFR of the Galaxy and
derived an average value of $1.9 \pm 0.4$ \msunyr. 
A more recent analysis of the same data using hierarchical Bayesian analysis found $1.65\pm 0.19$ \msunyr\ \citep{2015ApJ...806...96L}. 
We adopt this value while noting that systematic uncertainties, especially
assumptions about the initial mass function as discussed by
\citet{2015ApJ...806...96L},
likely allow uncertainties of about 50\%.

The most complete catalog of
structures identified from the most complete CO survey of the Galaxy
\citep{Dame01} is that of \citet{MD17}, hereafter referred to as MD. 
They were able to assign 98\% of the CO emission to 
8107 structures, to which they assigned sizes and velocity dispersions.
They found that much of the mass was in unbound structures, with the
virial parameter,
$\alphavir > 2$. 
Here, we follow convention in the literature of defining a virial parameter using the observed cloud effective radius, mass, and one-dimensional velocity dispersion as
\begin{equation}
\alphavir \equiv \frac{5 \sigma_{\rm 1d}^2 R}{G M}; 
\end{equation}
this ignores  magnetic terms, surface terms, internal inhomogeneity and stratification, and tidal gravity, all of which may be important at the factor of $\sim 2$ level \citep[e.g.][]{1992ApJ...399..551M,2015ApJ...809..154H,2020ApJ...898...52M,2021ApJ...911..128K}.  However, this definition has the virtue that it is relatively easily measured in observations, and it is also straightforward to control this parameter in the initial conditions of simulations.

Recently,
\citet{2021ApJ...920..126E} 
found that only 19\% of the mass in the MD structures was in gravitationlly
bound structures (i.e., having $\alphavir \le 2$).
If only bound clouds form stars at the free-fall rate, the
predicted SFR can be decreased to $46$ \msunyr, decreasing the discrepancy by a
factor of 6.5. This simple analysis assumed that the masses in the MD
catalog were correct and took a very simple (step-function) model of
how \epsff\ depends on \alphavir.

In this paper, we 
further investigate  implications for the predicted SFR 
of correcting cloud properties for
the known gradient in metallicity in the Milky Way and applying a formula
for \epsff\ based on simulations of clouds with various initial values of
\alphavir.

\section{Reconsidering the Mass and Related Properties of Galactic Molecular Structures}\label{aco}

Central to the determination of both the free-fall time and virial parameter
is an independent measure of the mass of the structure. For large-scale studies
of the molecular gas, the most common mass tracer is the line luminosity of
low-$J$ transitions of CO. The mass of the structure traced by CO, generally
called a molecular cloud, comes from $M_{\rm mol} = \alphaco \lco$, where $\lco$ is the CO luminosity and \alphaco\ (hereafter in units of \alphacounit) is the CO-to-H$_2$ mass conversion factor 
accounting for the mass contributions from associated helium and metals.
There is growing recognition that the conversion of CO luminosity into mass is unlikely to be the same in all environments
(see \citealt{2020ARA&A..58..157T} and references therein). 
The most obvious source of variation in the conversion factor is variation in metallicity. We use the symbol $Z$ to represent the metallicity relative to that of the solar neighborhood. 

Observers of other galaxies tend to correct \alphaco\ for $Z$, and
various formulas have been developed
\citep{2013ARA&A..51..207B, 2017MNRAS.470.4750A, 2020ARA&A..58..157T, 2020A&A...643A.141M}.
Many references 
adopt a conversion factor 
of the form
\begin{equation}\label{eq:Sun}
\alphaco = \alphacoz\ Z^{-a} , 
\end{equation}
with $a = 1.6$ a common choice
(e.g., \citealt{2020ApJ...901L...8S}) 
for nearby star-forming galaxies. The weakness of CO emission
in dwarf galaxies with very low metallicity 
has led to larger estimates for $a$:
2.0-2.8
\citep{2012AJ....143..138S},
or even
3.39
\citep{2020A&A...643A.141M}.

For the Milky Way, most studies have assumed that a commonly
adopted local calibration, 
taken to be $\alphaco = 4.35$, 
applies throughout the Galaxy. 
This luminosity-to-mass conversion is related to the commonly used relation
for \hh\ column density, $\nhh = \xco \ico$, with \ico\ the integral of main-beam temperature over velocity and the CO-to-H$_2$ column density conversion factor \xco\ (hereafter in units of \xcounit).
Based on local calibration, the most commonly used value is
$\xco = 2\ee{20}$.
However,
\citet{2020ApJ...898....3L} 
have suggested a correction factor that removes an apparent Galactic gradient in the mean surface density of clouds defined by CO:
\begin{equation}\label{eq:LD}
X_{\rm CO,LD} = \frac{83\ee{20}}{54.5 -3.7 (\rgal/{\rm kpc})}\,, 
\end{equation}
for Galactocentric radius $2$ kpc $< \rgal \leq 10$ kpc and $X_{\rm CO} = 6.0\ee{20}$ 
for $\rgal > 10$ kpc. While we do not find the arguments for this 
particular dependence
of \xco\ on \rgal\ compelling, the idea of variation of \alphaco\ in
the Milky Way would be consistent with practice for other galaxies.

Recently,
\citet{2020ApJ...903..142G}
have computed simulated CO emission from a set of kpc-scale numerical MHD simulations of the ISM,  and fitted the 
results to determine column density conversion factors \xco\ that depend on $Z$
and other observational parameters.  
\citet{2020ApJ...903..142G}
provide several possible formulas for \xco\ for the \jj10\ transition. 
The first is a simple formula involving only $Z$ (``Gong1a'')
\begin{equation}\label{eq:Gong1a}
X_{\rm CO,Gong1a} = 1.4\ee{20} Z^{-0.80},
\end{equation}
which has a substantially less steep dependence on $Z$ than 
is often adopted in extragalactic observations.
A similar dependence ($Z^{-0.7}$) was found in simulations including
time dependent effects by
\citet{2022arXiv220103885H}.
Improved comparison to the actual values of \nhh\ in the 
\citet{2020ApJ...903..142G}
simulations is
provided by a formula that also accounts for the intensity (\ico) 
of the emission and the spatial resolution of the survey (\rbeam). 
This expression 
(``Gong4a'') is 
\begin{eqnarray}\label{eq:Gong4a}
X_{\rm CO,Gong4a} = 6.1\ee{20} Z^{-0.80}  \rbeam^{-0.25} \times \nonumber  \\
 \ico^{-0.54 + 0.19\log \rbeam},
\end{eqnarray}
where $r_{\rm beam}$ is in pc and \ico\ is in \kkms.
This formula is recommended only if the effective resolution is $\lesssim 100$ pc.
The resolution of the 
\citet{Dame01}
survey is about 8\farcm5, translating to $\rbeam = 2.5 (d /{\rm kpc})$ pc, which
reaches 100 pc only for $d = 40$ kpc; the formula is
suitable for the Milky Way.

We will test each of the above formulations for converting CO emission to molecular mass. To obtain \ico\ from the values
tabulated as ``WCO" in the machine-readable table in \citet{MD17}, we divide by the number of pixels, also given in that table, in accordance
with Equation 13 of \citet{MD17}. The value of \ico\ is then an average over
the structure.

To convert these column density factors to mass 
conversion factors, we use
\begin{eqnarray}
\sigmam  & = & \muhh m_{\rm H} \nhh \,,\; {\rm or} \nonumber  \\
\sigmam (\msunpc) 
& = & 8.015\ee{-21} \muhh \nhh (\cmc),
\end{eqnarray}
where
$\nhh = \xco \ico$ for the region identified by CO emission, and $\Sigma_{\rm mol}$ includes helium and metals.
Using the values for $M_{\rm X}/M_{\rm H}$ for the latest
proto-solar abundances from 
\citet{2021A&A...653A.141A}, the mean molecular weight
$\muhh = 2.809$, $\sigmam (\msunpc) = 2.251\ee{-20} \nhh (\cmc)$, and
\begin{equation}
  \alpha_{\rm CO} = 4.50 \left( \frac{X_{\rm CO}}{2.0 \times 10^{20}} \right) \, ,
\end{equation}
This conversion is about 3\% higher than the usual choice of
$\alphaco = (4.35/2\ee{20})\xco$.

To convert the $Z$ dependence of \alphaco\ into a dependence on \rgal, we
use measured radial gradients in $Z$.
There is compelling evidence for a gradient in $Z$ in the Milky Way,
with consistent evidence from pulsation of Cepheids
(e.g., \citealt{2018A&A...618A.160L}),
to \hii\ region electron temperatures from radio recombination lines
(e.g., \citealt{2019ApJ...887..114W}),
to direct determinations of abundances in \hii\ regions from 
optical spectral lines.
We use  abundances measured in \hii\ regions because these represent abundances within the last 5-10 Myr, so most relevant to
current conditions in molecular clouds
\citep{2000MNRAS.311..329D}.
The most recent measurements find a steady decrease in metal abundances from
4 to 17 kpc in \rgal\ \citep{2018PASP..130k4301W, 2020MNRAS.496.1051A, 2021MNRAS.502..225A, 2022MNRAS.510.4436M}.
The last reference combines all the data with EDR3 distances from 
Gaia to provide gradients for both oxygen and carbon, both with and without corrections
for temperature inhomogeneities.

The two constituents of CO appear
to have different gradients and the O/H gradient depends
slightly on assumptions about temperature inhomogeneities. 
Because the various formulas
for varying \alphaco\ assumed that all abundances (including dust) scale together, characterized by $Z$, with
$Z = 1$ at the solar circle, we need
to choose one gradient.
Which relation to use is not entirely clear. Because carbon is less abundant
than oxygen, the CO abundance is likely limited by carbon rather than oxygen.
The dust opacity in the ultraviolet, important to the shielding in the 
simulations, is contributed roughly equally by carbonaceous and silicate dust
(Fig. 23.11 of \citealt{2011piim.book.....D}).
Carbon has the additional advantage that corrections for temperature variations
do not affect the gradient in C, unlike the case of O.
On the other hand, the oxygen abundance has been determined in more \hii\
regions over a wider range of \rgal\ (5-17 kpc) versus 6-12 kpc for carbon.
Because no variation of \alphaco\ with $Z$ has been usually
assumed for Galactic studies, we make a ``conservative"  choice
of the smallest gradient: that for O/H without correction for
temperature inhomogeneities. This gradient is $-0.044$ dex/kpc
\citep{2022MNRAS.510.4436M}. 
The gradient measured from double-mode pulsating Cepheids 
($-0.045 \pm 0.007$ dex/kpc) agrees with this choice
\citep{2018A&A...618A.160L}. 
We consider the effect of other choices in
\S \ref{caveats}. 

\begin{figure}
\center
\includegraphics[scale=0.4]{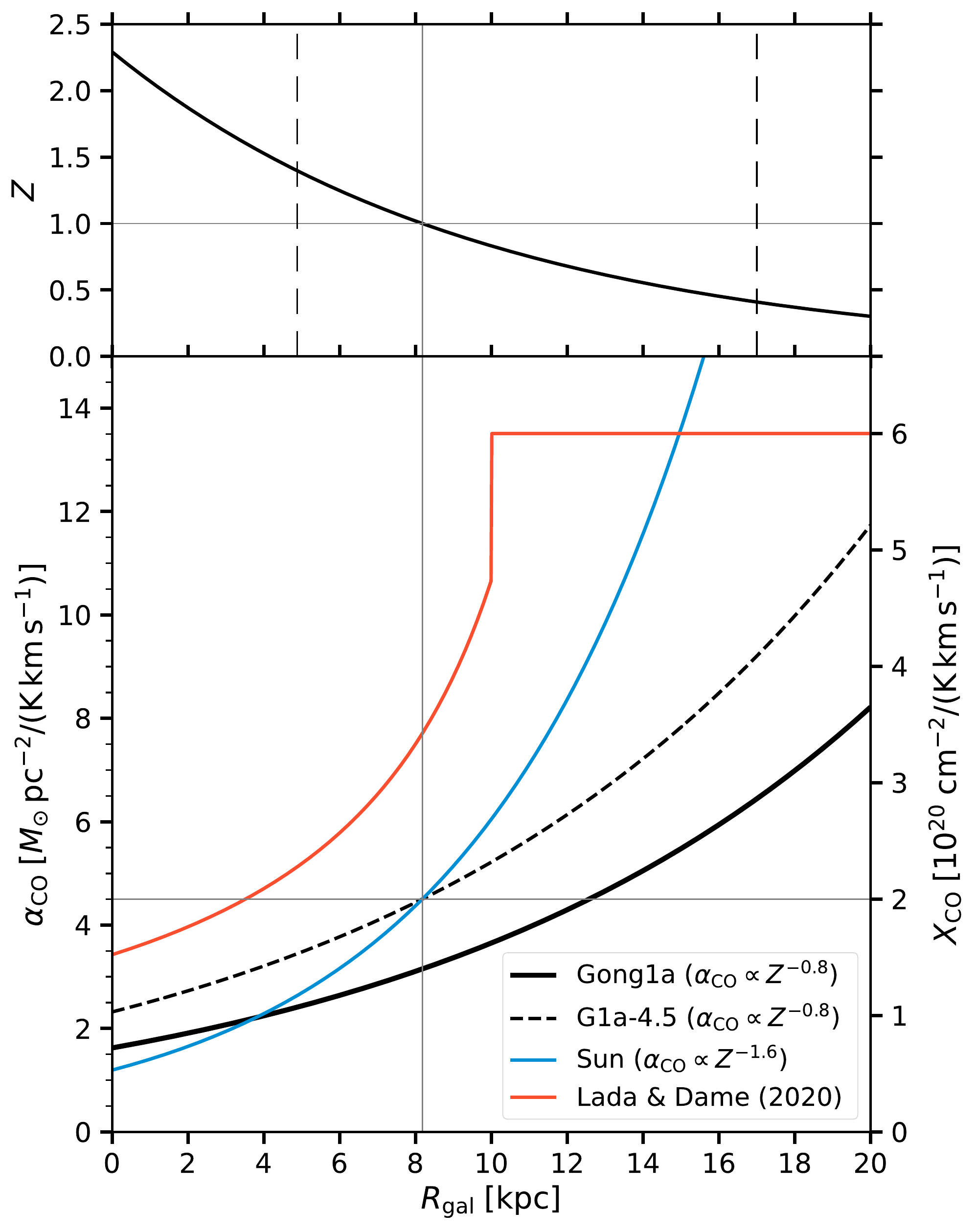}
\caption{
(Top) Assumed variation of $Z$ with \rgal\ based on the  
radial abundance gradient of  
oxygen in \hii\ regions: 
$c= -0.044$ dex/kpc
\citep{2022MNRAS.510.4436M}. The dashed vertical lines represent the range of $R_{\rm gal}$ for which the abundance gradient is measured. (Bottom) Various models of \alphaco\ versus \rgal. The grey vertical line marks the solar circle, $R_{\rm gal,\odot} = 8.178$ kpc \citep{2019A&A...625L..10G}, and the grey horizontal
lines indicate the usual assumptions for the solar neighborhood, $Z=1$ and $\alpha_{\rm CO,0}$ updated to 4.50 \alphacounit.}
\label{acomodels}
\end{figure}

To translate these gradients into $Z$, the metallicity used by 
\citet{2020ApJ...903..142G}, we normalize to $Z = 1$ at the distance from the center of the Galaxy to 
the solar neighborhood of $\rgalsun = 8.178 \pm 0.013_{\rm stat.} \pm 0.022_{\rm sys.} $ kpc 
\citep{2019A&A...625L..10G}. 
We assume 
\begin{equation}
Z = 10^{c \left( \rgal - \rgalsun \right)},
\end{equation}
with $c = - 0.044$ dex/kpc. This is plotted in the top panel of Figure~\ref{acomodels}.

We plot the relation labeled  Gong1a 
\citep{2020ApJ...903..142G} 
 with the adopted $Z$ gradient, along with that used by 
\citet{2020ApJ...901L...8S}
for other galaxies (but with \alphacoz\ rescaled from 4.35 to 4.5),
and the relation from
\citet{2020ApJ...898....3L} in the bottom panel of Figure \ref{acomodels}.
There are quite large differences, especially at large radii.

The Gong1a formula predicts a local
value of $\xcoz = 1.4\ee{20}$, smaller
than the usually accepted $\xcoz = 2.0\ee{20}$
\citep{2010ApJ...721..686P,2013ARA&A..51..207B}.
The lower value was the best fit to all the simulations, but
their R4 and R8 simulations, more representative of the local
molecular gas, favored values closer to $\xcoz = 2.0\ee{20}$,
as can be seen in Fig. 6 and 9 of
\citet{2020ApJ...903..142G}.
The median and mean \ico\ in the MD catalog are 3.8 and 7.2 \kkms,
values for which Gong1a underestimates \xcoz\ (Fig. 13 of
\citealt{2020ApJ...903..142G}).
For these reasons, we also include a modified Gong1a model
(labeled G1a-4.5),  
with $\xco = 2.0\ee{20}Z^{-0.80}$,
leading to $\alphaco = 4.50 Z^{-0.80}$. 

\section{The Dependence of \epsff\ on \alphavir }\label{avir}

A very simple treatment of the star formation efficiency was adopted by \citet{2021ApJ...920..126E}, who assumed a step function with a transition from 1 to 0 at $\alphavir=2$. 
More realistically, theory and simulations predict that for a given molecular cloud mass and size, the star formation rate would be systematically lower if the turbulence amplitude is higher, corresponding to  higher virial parameter.  

Based on a set of self-gravitating driven-turbulence simulations, 
\citet{2012ApJ...759L..27P}
proposed that the efficiency per free fall time, \epsff, depends exponentially on the ratio of free-fall time to dynamical time: $\epsff = \epscs \exp(-C t_{\rm ff}/t_{\rm dyn})$, where \epscs\ is the core-to-star efficiency and $C$ is determined from simulations that resolve cloud-to-core scales (but not core-to-star scales). 
Their simulations were for a periodic box; defining 
$t_{\rm dyn} = r/\sigma_{\rm 1d}$
for $r$ half the box length, their proposed star formation law corresponds to $C=2.77$.
For a spherical cloud, the ratio of free-fall and dynamical times is related to the virial parameter by $t_{\rm ff}/t_{\rm dyn} = \pi (\alphavir/40)^{1/2}$, so we can rewrite the proposed expression for 
the cloud-to-core efficiency per free-fall time as
\begin{equation}\label{koeqn}
\epsilon_{\rm ff,cc} =  \exp(-b \alphavir^{1/2})
\end{equation}
where $b = C \pi/\sqrt{40}$.
The \citet{2012ApJ...759L..27P} fit to their simulations corresponds to $b=1.38$
(indicated by PN in figures).

\citet[][hereafter KOF]{2021ApJ...911..128K} conducted a set of isolated, turbulent-cloud
simulations with UV radiation feedback for varying initial virial parameter \alphavirz, and fit their results for the SFR to find a best fitting value of $C = 4.06$, or $b = 2.02$ in Equation~\ref{koeqn}, indicating a somewhat stronger dependence than originally suggested by 
\citet{2012ApJ...759L..27P}. 
The KOF simulations considered clouds of fixed initial mass and radius $(M_0,R_0) = (10^5\,\msun, 20\pc)$. We have run additional simulations of more massive, low-density clouds $(M_0,R_0) = (10^6\,\msun, 60\pc)$ with $\alphavirz =1$, $2$, $5$, and $10$. Since these clouds are relatively long-lived compared to UV-emitting stars, effects of supernova explosions are also included following the procedures of \citet{2017ApJ...846..133K}. We found that with these extensions, $b \approx 2$ still well describes the decreasing trend of measured efficiency per free-fall time with $\alphavirz$.

Because the simulations of PN and KOF were not resolved at the core to star level, their efficiency was really the core formation efficiency and the core-to-star efficiency (\epscs) must come from other considerations.
\citet{2012ApJ...759L..27P}
chose $\epscs = 0.5$ to account for mass lost through jets and winds.
Comparison of the core mass function to the stellar mass function 
\citep{2007A&A...462L..17A, 2008ApJ...684.1240E, 2015A&A...584A..91K}
and simulations of envelope clearing by winds
\citep{2010ApJ...710..470D}
suggest $\epscs = 0.20$ to $0.40$; 
we adopt $\epscs = 0.30$. 
Using \epscs\ implies that the mass removed from a core by outflows is no longer available for star formation in the cloud. Outflow velocities typically exceed the cloud's escape velocity and breakouts from the cloud are observed (e.g.,
\citealt{2004ApJS..154..352N}).  In comparing to the observed value, \epsffobs, we use the product of 
$\epsilon_{\rm ff,cc}$ from simulations and \epscs.

\begin{figure}
\center
\includegraphics[scale=0.55]{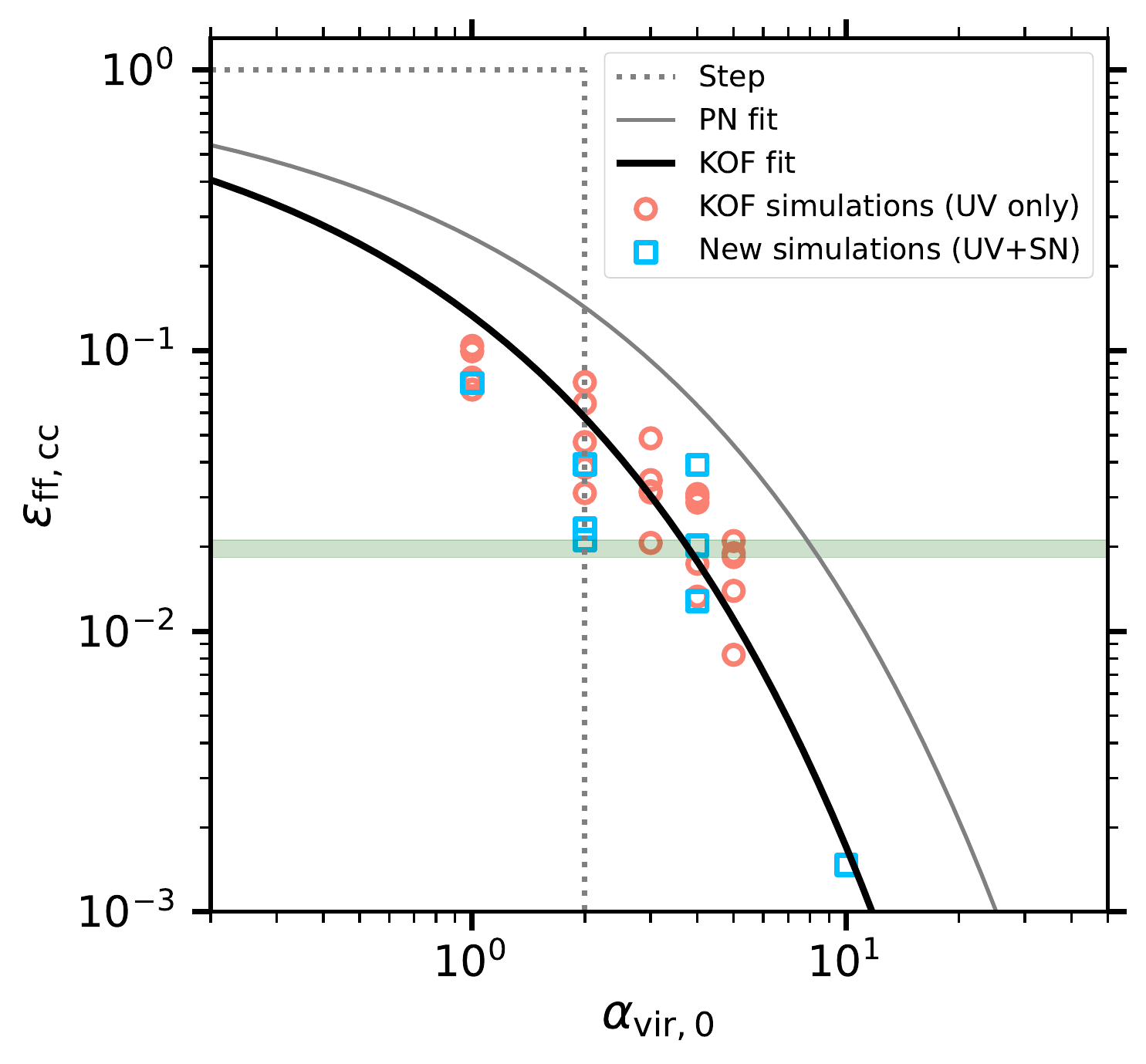}
\caption{Assumed relation between cloud-to-core efficiency per free-fall time and the virial parameter of a molecular cloud. The thick black line shows our standard choice with $b=2.02$ (see Equation~\eqref{koeqn}), which is a fit to
the $\alpha$-series simulations of KOF (open circles)
with UV radiation feedback.
The squares show the results of new simulations of more massive clouds 
including both UV radiation and supernovae feedback. The thin grey line shows the relation proposed by
\citet{2012ApJ...759L..27P} with $b=1.38$. The dotted line shows the simple step-function relation adopted by \citet{2021ApJ...920..126E}. The green shaded region indicates the observational constraint $ \epsilon_{\rm ff, obs} / \epsilon_{\rm cs} = ({\rm SFR}_{\rm obs}/{\rm SFR}_{\rm ff})/\epsilon_{\rm cs}$ for ${\rm SFR}_{\rm ff}/{\rm SFR}_{\rm obs} = 158$--$182$ and $\epsilon_{\rm cs}=0.3$.}
\label{epsff}
\end{figure}

In Figure~\ref{epsff}, we show results from simulations for $\epsilon_{\rm ff,cc}$ as a function of the initial virial parameter (symbols), and the fit to simulation results (lines), in comparison to the mean observed $ \epsffobs/\epsilon_{\rm cs}$ (shaded region) with $\epscs = 0.30$.

\section{Comparison of Model and Observations}\label{thtest}

We follow 
\citet{2021ApJ...920..126E}
in using all the entries in the \citet{MD17} catalog that satisfy the conditions
$M_{\rm mol} > 1$ \msun, $R_{\rm gal} < 30$ kpc,  and $\alphavir < 100$. 
There is very little mass outside these restrictions but they remove
some outliers. 

We calculated the free-fall star formation rate as 
\begin{equation}
    \sfrff = \sum_{\rm cl} M_{\rm mol}/t_{\rm ff} \equiv M_{\rm mol,tot}/\langle \tff \rangle\,,
\end{equation}
where 
$t_{\rm ff} = [3\pi/(32G \rho)]^{1/2}$ with $\rho = 3M_{\rm mol}/(4\pi R^3)$ and
the summation is taken over the cloud sample defined above and $\langle t_{\rm ff}\rangle$ is the mass-weighted harmonic mean of the free-fall time. The theoretical prediction allowing for the variation of 
$\epsff$ is
\begin{equation}\label{eq:SFRpred}
\sfrth = \sum_{\rm cl} 
\epsff 
M_{\rm mol}/\tff \equiv 
\langle \epsff \rangle \sfrff\,,
\end{equation}
where $\langle \epsff \rangle$ is the average 
$\epsff$
weighted by the free-fall star formation rate.
For simulations that do not resolve the core-to-star efficiency, $\epsff = \epscs\epsilon_{\rm ff,cc} $.  To match observations,  $\langle\epsilon_{\rm ff,cc}\rangle$ must equal $\epsffobs/\epscs = 0.006/0.3 = 0.02$.

First, we explore the effects of a varying \alphaco, while retaining
the simplistic assumption by
\citet{2021ApJ...920..126E}
of a step function for \epsff:
$\epsff = 1$ for $\alphavir \leq 2$ and 
$\epsff = 0$ for $\alphavir > 2$
(with \epscs\ as unity).
For this test, we consider (1) a constant $\alphaco=4.5$; (2) the formula based on Equation \ref{eq:LD} labeled Lada \& Dame in Figure~\ref{acomodels} and LD in Table~\ref{tabresults}; (3) the expression in 
Equation~\ref{eq:Sun} with $a = 1.6$ labeled Sun; (4) the expression in 
Equation~\ref{eq:Gong1a}
labeled Gong1a; (5) the expression
in Equation~\ref{eq:Gong4a}
labeled Gong4a.

Entries 1-5 in Table \ref{tabresults} for $\thdobsstep = {\rm SFR}_{\rm step}/{\rm SFR}_{\rm obs}$ show that 
these predictions are all better than allowing all clouds to form
stars at the free-fall rate, which has $\thdobsff = 158$.
However, the LD formulation overproduces stars by the largest factor
($\thdobsstep = 73$) while the Gong1a formula minimized the problem, with
$\thdobsstep = 3.4$. 
The trends are partially explained by the column showing the total
molecular mass. The decreasing \alphaco\ in the inner Galaxy decreases
the total mass of molecular gas to 
1.0\ee9 \msun\ 
for the Gong1a formula,
while the LD formula actually increases the Galaxy's molecular mass to 
2.3\ee9 \msun. 
An additional contributing factor is shown by column 4 with
\mean{\tff}, which is longer for the two Gong entries and the Sun entry. 
The full effect of \alphacoz, including the higher \alphavir\ 
is indicated by column 7 with \mean{\epsff}.
 Clearly, taking account of the variation of \alphaco\
has a very large effect on the predicted star formation rate, but it
does not by itself solve the problem entirely.

\begin{deluxetable}{ccccccc}
\tabletypesize{\footnotesize}
\tablewidth{0pt}\tablecaption{Summary of Results}
\tablehead{
\colhead{$\alpha_{\rm CO}$} &
\colhead{SFE} &
\colhead{$M_{\rm mol,tot}$} &
\colhead{$\langle t_{\rm ff} \rangle$} &
\colhead{${\rm SFR}_{\rm th}$} &
\colhead{$Q_{\rm th}$} &
\colhead{$\langle \epsilon_{\rm ff} \rangle$} \\
\colhead{} &
\colhead{} &
\colhead{($10^9\,$\msun)} &
\colhead{(Myr)} &
\colhead{(\msunyr)} &
\colhead{} &
\colhead{(\%)}\\
\tableline
\colhead{(1)} &
\colhead{(2)} &
\colhead{(3)} &
\colhead{(4)} &
\colhead{(5)} &
\colhead{(6)} &
\colhead{(7)}
}
\startdata
4.50 & Step & 1.7 & 7.4 & 52.7 & 32 & 23  \\
LD & Step & 2.3 & 7.5 & 120 & 73 & 39  \\ 
Sun & Step & 1.3 & 10.4 & 33.2 & 20 & 26  \\
Gong1a & Step & 1.0 & 10.9 & 5.67 & 3.4 & 6.3  \\
Gong4a & Step & 1.1 & 12.3 & 8.47 & 5.1 & 9.6  \\
\tableline
4.50 & KOF & 1.7 & 7.4 & 2.66 & 1.6 & 1.2  \\
LD & KOF & 2.3 & 7.5 & 5.93 & 3.6 & 1.9  \\
Sun & KOF & 1.3 & 10.4 & 1.99 & 1.2 & 1.5  \\
Gong1a & KOF & 1.0 & 10.9 & 0.50 & 0.30 & 0.55  \\
Gong4a & KOF & 1.1 & 12.3 & 0.57 & 0.35 & 0.65  \\
\tableline
G1a-4.5\tablenotemark{a} & KOF & 1.4 & 9.2 & 1.46 & 0.89 & 0.95  \\
\enddata
\tablecomments{Column (1) gives the model treatment adopted for $\alpha_\mathrm{CO}$ (see \autoref{obs}). Column (2) gives the model treatment adopted for $\epsff$ to set the efficiency of star formation (see \autoref{avir}). Columns (3)--(5) give the total cloud mass, the mean value of $\tff$, and the total predicted SFR. Column (6) gives the ratio of predicted to observed SFR. Column (7) gives the mean efficiency per free-fall time from \autoref{eq:SFRpred}.}
\tablenotetext{a}{Gong1a with $\alpha_{\rm CO,0} = 4.50$}
\label{tabresults}
\end{deluxetable}

Second, we explore the effects of including the dependence of $\epsff$ on $\alphavir$  from
Equation~\ref{koeqn}.  We 
adopt the KOF model 
setting $\epsff = \epscs \epsilon_{\rm ff,cc}$
and $\epsilon_{\rm ff,cc}=\exp(-2.02 \alpha_{\rm vir}^{1/2})$
with $\alphavir$ equal to the observed 
value in each cloud, and $\epscs = 0.30$.
The results of this exercise are shown in  
entries 6-10 of Table \ref{tabresults}.
For constant $\alphaco=4.50$, 
with $\epscs = 0.30$, 
we obtain a predicted star formation rate of 2.66 \msunyr, nearly consistent with the observational constraint. 
Combining the LD conversion factors with the KOF 
prediction for $\epsilon_{\rm ff,cc}$ predicts too high a SFR.  
If we apply either of the \citet{2020ApJ...903..142G} formulas for \alphaco, 
the star formation rate is actually slightly
under-predicted. Entry 11 in
Table \ref{tabresults} (denoted G1a-4.5) uses the scaled up Gong1a
formula; the resulting 
$\sfrth = 1.46$ \msunyr, 
the closest to the observed values.
Given uncertainties in quantities like \epscs, and the \rgal\ 
dependence of \alphaco, the agreement with observations is reasonable for
several combinations.

\begin{figure}
\epsscale{1.2}\plotone{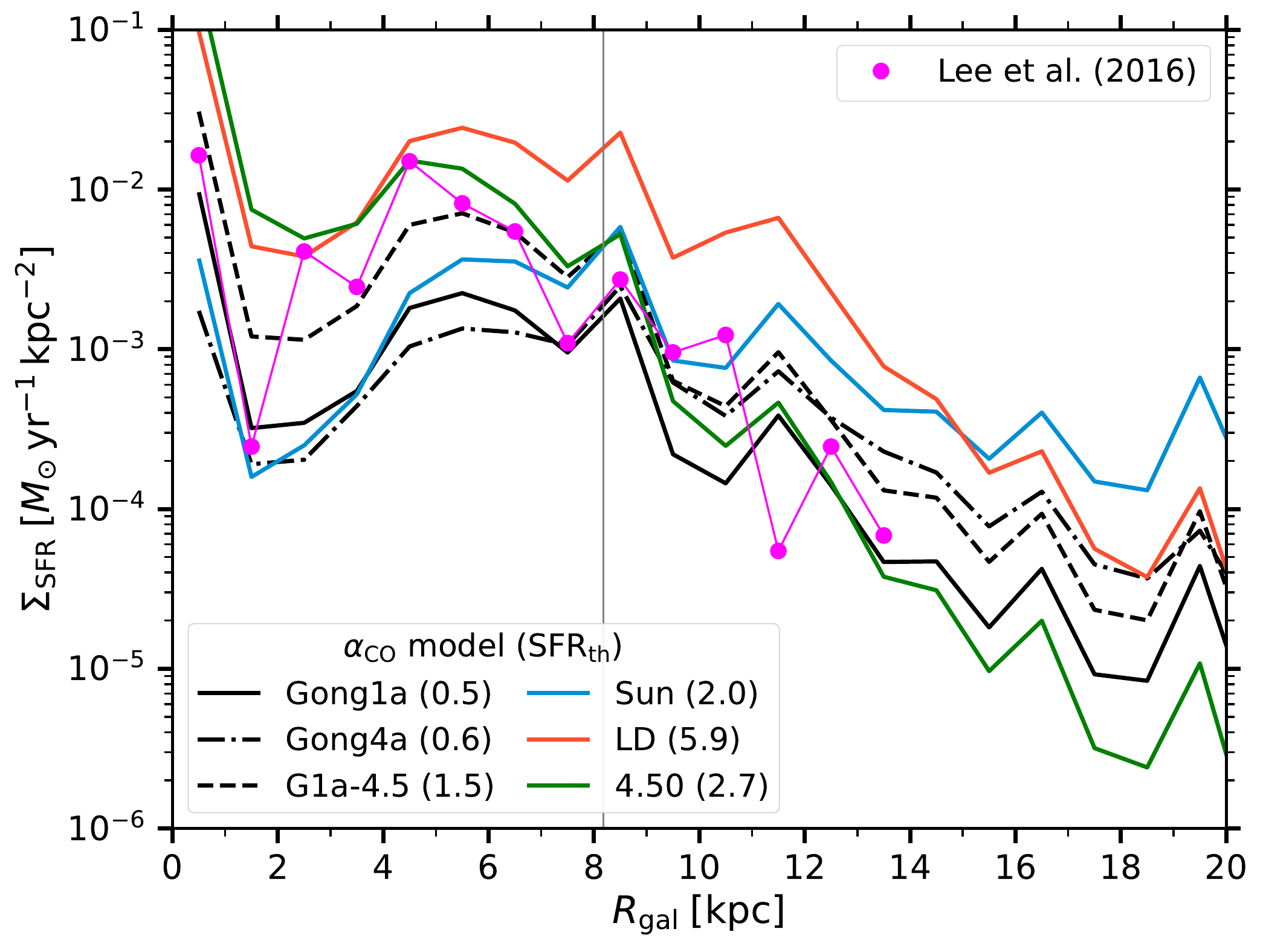}
\caption{The log of the star formation rate surface density is plotted versus \rgal. 
The magenta circles are observational estimates taken from 
Fig. 9 of
\citet{2016ApJ...833..229L},
with the total star formation rate scaled to 1.65 \msunyr.
The other lines are those predicted by different models of the dependence of \alphaco\ on \rgal, as indicated by the label. The numbers in parentheses indicate the predicted total star formation rate \sfrth\ in \msunyr (see also Table~\ref{tabresults}). All use the KOF formula for 
$\epsilon_{\rm ff,cc}$
with $b=2.02$.
}\label{sfrvsrgal}
\end{figure}

We can also compare predicted and observed distributions of the star formation rate over \rgal.
In Figure \ref{sfrvsrgal} we show the predictions of the six models for \alphaco,
all using the KOF model for 
$\epsilon_{\rm ff,cc}$
to predict the surface density of star formation rate $\Sigma_{\rm SFR}$ in bins of \rgal.
The points representing the observations are taken from
\citet{2016ApJ...833..229L},
but scaled up so that the total star formation rate is 1.65 \msunyr.
These are based on 191 associations of the strongest {\it WMAP} 
free-free emission with molecular clouds in the MD survey.

The model of a constant \alphaco\ clearly exceeds the observations
in the inner Galaxy, but performs well at large \rgal.
The LD formula for \alphaco\ over-predicts $\Sigma_{\rm SFR}$ 
over almost all of the Galaxy.
The Sun formula, with its very strong dependence on $Z$, 
underpredicts the star formation rate in the inner Galaxy and
predicts substantial star formation in the far outer Galaxy.
The Gong1a and Gong4a formulas predict a distribution that is too
flat in \rgal, with too little star formation from 2--8 kpc. 
The G1a-4.5 model does quite well, but may exceed the observations beyond
about 11 kpc. 

There is a well-known issue
that star formation in the CMZ (inner few 100 pc) is far less than predicted by
the most commonly used models
\citep{2017MNRAS.469.2263B}.
We do not specifically consider the CMZ, but
the G1a-4.5 model predicts a star formation rate close to 
the observed value at $\rgal = 0.5$ kpc.
The anomalously low SFR in the CMZ may result in part from using
a constant \alphaco.

\begin{figure}
\center
\includegraphics[scale=0.45]{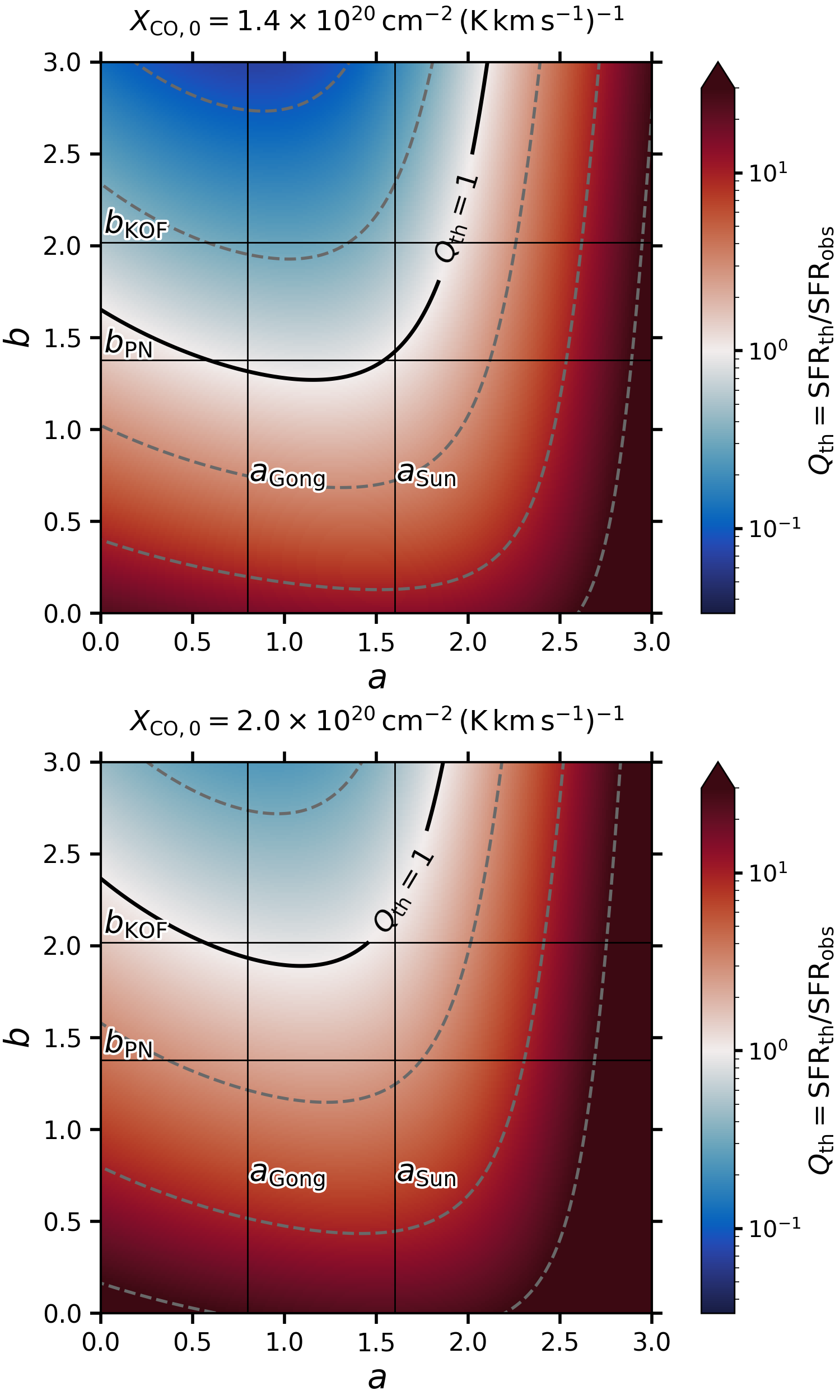}
\caption{Plots of \thdobs\ in the plane of $a$, $b$, where $b$ is the coefficient 
in the formula for \epsff\ and $a$ is the exponent in the formula for 
the $Z$ dependence of \alphaco. The upper panel assumes $\xcoz = 1.4\ee{20}$
\xcounit\ 
for the solar neighborhood and the lower panel assumes $\xcoz = 2.0\ee{20}$
\xcounit.
The locus of parameters that produce $\thdobs = 1$ (theory matches
observation) is shown as black solid curves, while grey dashed curves show the
region with a discrepancy by a factor of 3, 10, and 30.
The vertical thin lines show the values of $a$ recommended by 
\citet{2020ApJ...903..142G}
(Gong1a)
and used by 
\citet{2020ApJ...901L...8S}
; the horizontal thin lines show the values of $b$ from
\citet{2021ApJ...911..128K} and
\citet{2012ApJ...759L..27P}
}
\label{twoD}
\end{figure}

If we use the simpler versions of \alphaco\ that depend on only $Z$,
there are essentially two main parameters to consider: the coefficient $b$ in the exponent of \autoref{koeqn}; and the exponent $a$ 
in the $Z$ dependence of \alphaco\ (\autoref{eq:Sun})
Figure \ref{twoD} shows the color-coded value of $\thdobs$ in the $a$--$b$ plane; all models assume $\epscs=0.3$.
The upper panel is for $\xcoz = 1.4\ee{20}$, 
as \citet{2020ApJ...903..142G} found, while the lower panel is for
$\xcoz = 2.0\ee{20}$, 
which corresponds to adjusting
the Gong1a model to the usually assumed local \xcoz.
We note that the predicted SFR depends sensitively on the scaling of \xcoz. For a cloud with $\alpha_{\rm vir}=2$,
for example, scaling \xcoz\ by a factor of 2 (0.5) results in a factor of 6.5 increase (10 decrease) in the predicted star formation rate.
The strong dependence arises because mass, free-fall time, and
virial parameter all depend on the conversion factor.
Nevertheless, the plots indicate the range of values for
both parameters that fit the observational value within uncertainties.

The color gradients in Figure \ref{twoD} clearly demonstrate a greater sensitivity to $b$ in \autoref{koeqn} than to $a$ in \autoref{eq:Sun} for $a \lesssim 2$.
Values of $a > 2$ rapidly degrade the match to the total observed star formation rate. 
In particular, the large values inferred from studies of dwarf galaxies
\citep[e.g.,][]{2012AJ....143..138S,2020A&A...643A.141M}
cannot work for the Milky Way. They also
lead to too much gas mass and star formation in the outer Galaxy and conflict with the dependence on \rgal\ \citep{2016ApJ...833..229L}.
Quantitatively, we can separate the
effects by comparing values in Table \ref{tabresults}. Comparing entries 1 and 5, \thdobs\ decreases by a factor of 6 when Gong4a is used versus a constant 
$\alphaco = 4.5$. Comparing entries 1 and 6, \thdobs\ decreases by a factor of
20 when the KOF formula is used instead of the step function.
In contrast, the $Z$ dependence of \alphaco\ is most notable in the
dependence on \rgal\ (Figure \ref{sfrvsrgal}). 
This distinction will allow better separation of the
two effects with improved knowledge of how the Galaxy's SFR surface density depends on \rgal.

To conclude, we can predict within reasonable accuracy the SFR of the 
Galaxy by correcting molecular gas properties for the Galaxy's metallicity gradient and
including the dependence of \epsff\ on \alphavir\ found in calculations of the star formation
efficiency per free-fall time. This is a major achievement for the theory.

\section{Discussion}\label{disc}

\subsection{Properties of Structures as a Function of Mass}\label{cloudprops}

\begin{figure}
\center
\includegraphics[scale=0.25]{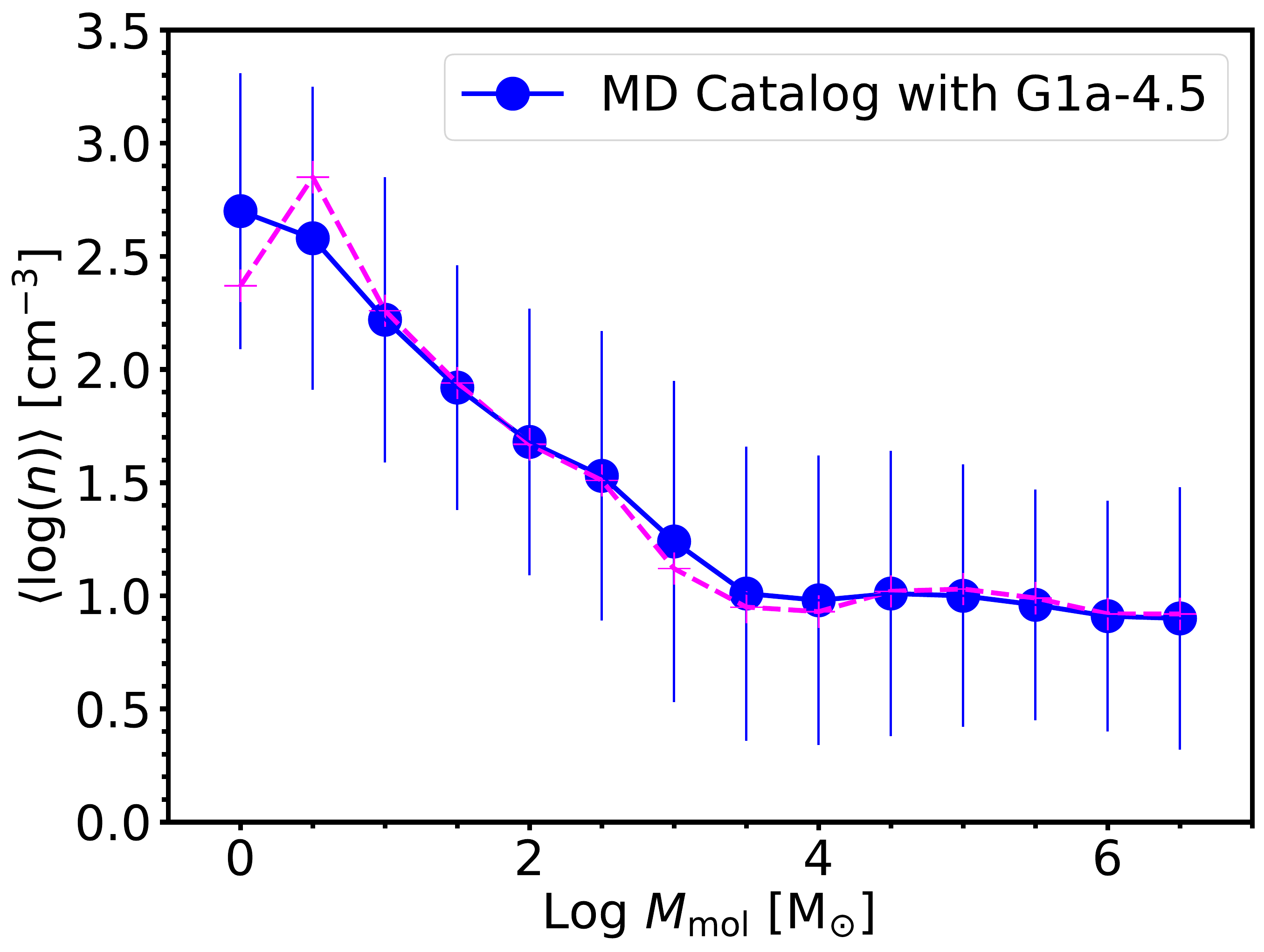}
\includegraphics[scale=0.25]{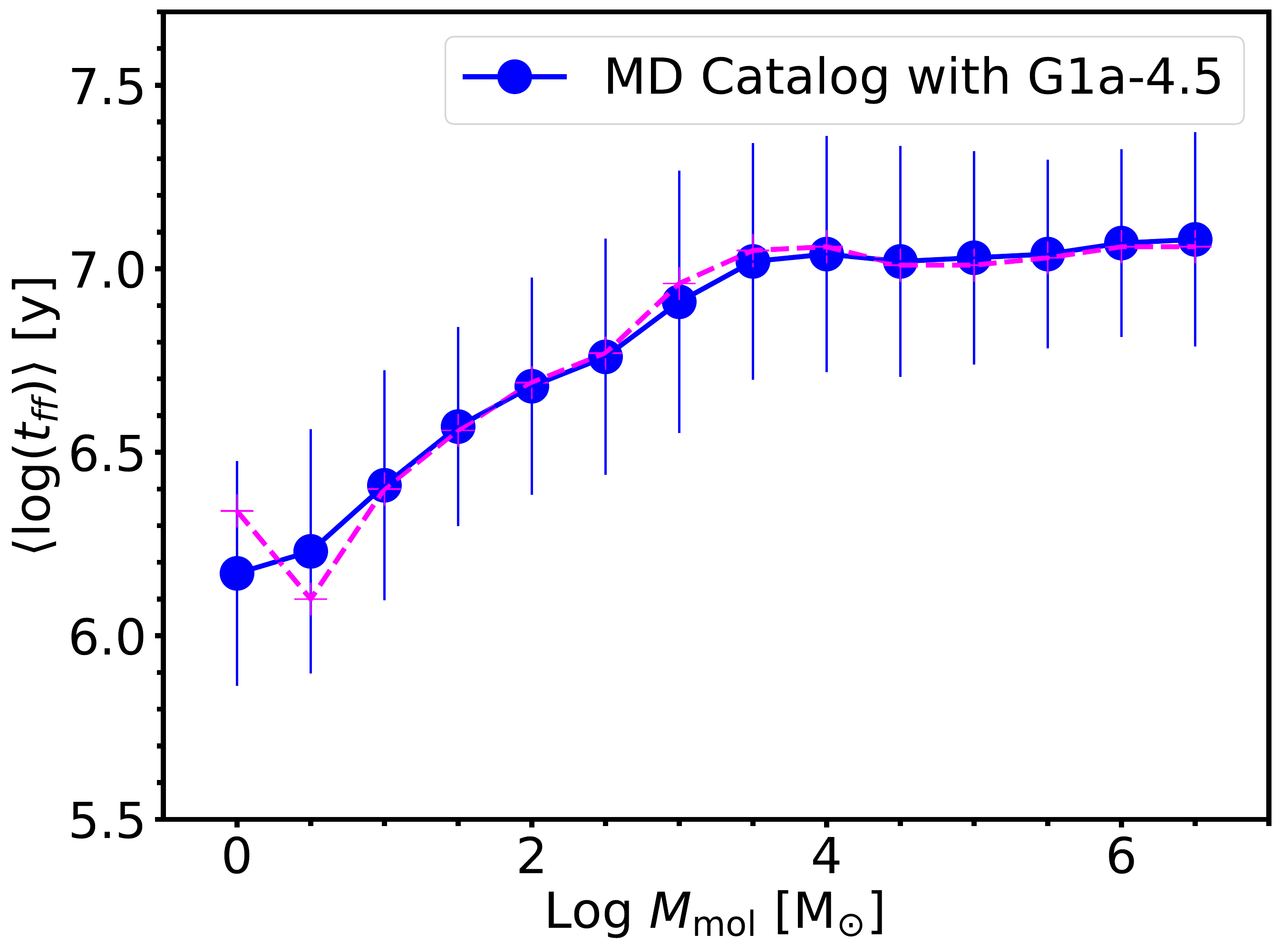}
\caption{(Top) The mean and standard deviation of $\log n$ (blue) in bins of 0.5 in
  $\log M_{\rm mol}$ for the cloud catalog of \citet{MD17} .
  The median is plotted in magenta. (Bottom) The log of the
  free-fall time versus $\log M_{\rm mol}$. The masses were computed using the G1a-4.5 formula to correct
  \alphaco\ for $Z$.
 }
\label{xvsM}
\end{figure}

One reason that the predicted star formation rate is less than the canonical 
calculation 
in \S \ref{s:intro}
when we let \alphaco\ vary with \rgal\ 
is that the density is lower and the free-fall time higher in the 
more massive clouds within the mass distribution, which are the ones with lower \alphavir\  \citep{2021ApJ...920..126E}.
The median, mean, and standard deviation of the total 
volume density of all particles, $n = n_{\rm H_{2}} + n_{\rm He} + \ldots$, and
the same statistics for the free-fall time are plotted versus $\log M$ in 
Figure \ref{xvsM}. These plots assume the G1a-4.5 formula for \alphaco.
The massive clouds have mean densities, $n \sim 10$ \cmv, much lower than are usually assumed.
The observational detection limits can be met even at these low densities,
as the effective density to produce a CO \jj10\ line of 1 \kkms\ is 15 \cmv\
\citep{2021ApJ...920..126E}.  
The free-fall time increases strongly with \mco\ up to 
about 3000 \msun, where it plateaus at about 10 Myr, substantially longer
than is usually assumed for a density of 100 \cmv, at about 3 Myr.

\subsection{Distributions of Mass and Star Formation Rate}

\begin{figure}
\center
\includegraphics[scale=0.5]{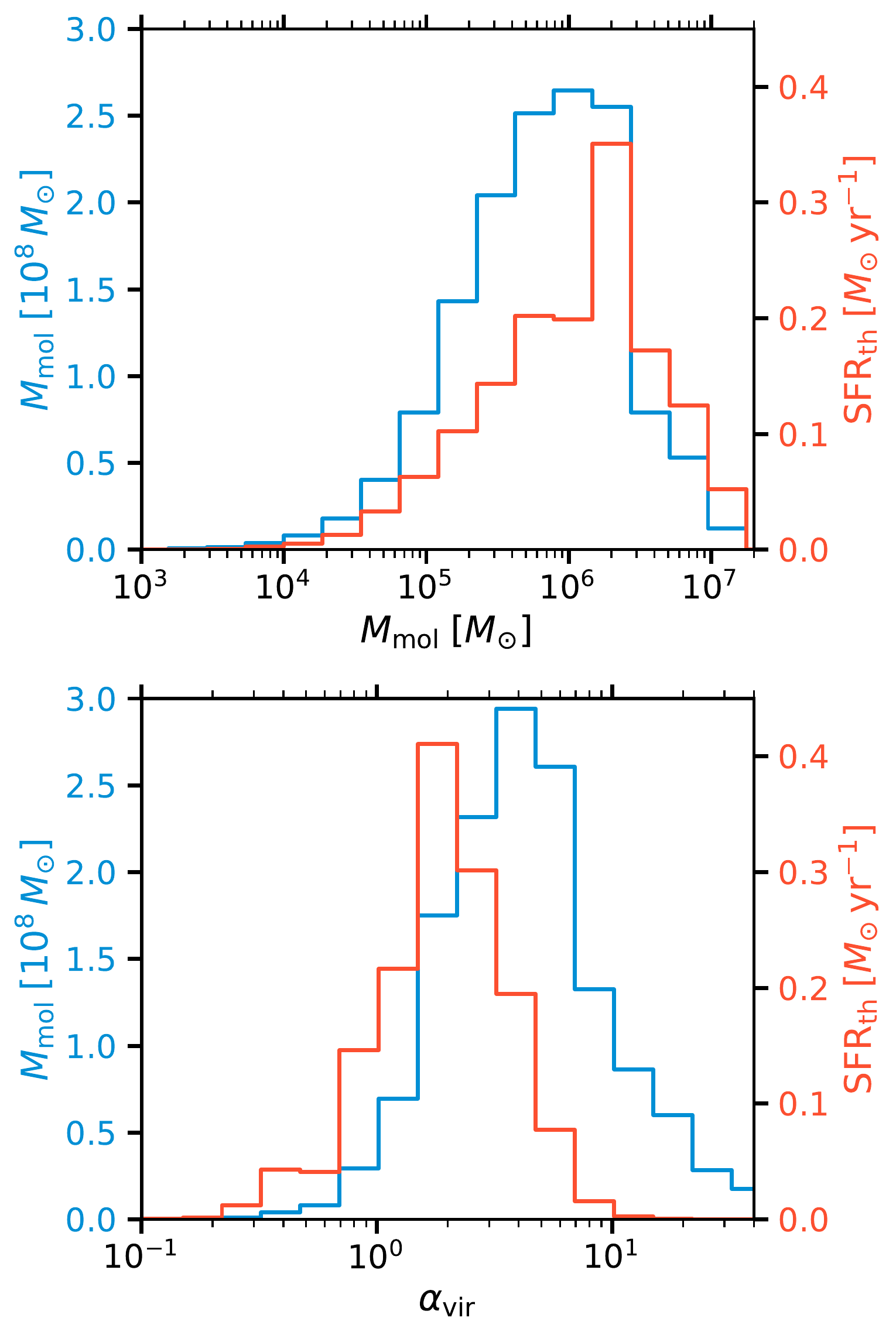}
\caption{The total mass (red) and predicted star formation rate (blue) of molecular clouds in bins of $M_{\rm mol}$ bin (top) and $\alpha_{\rm vir}$ (bottom) bin, adopting $\alphaco = 4.50 \times Z^{-0.8}$ (G1a-4.5) and $b_{\rm KOF}=2.02$. The total predicted star formation rate is 
$1.46$ \msunyr 
and the total molecular mass is 
$1.4\times 10^9$ \msun.}%
\label{histplot}
\end{figure}

\autoref{histplot} shows distributions of molecular mass and star formation rate in each mass (top) and $\alphavir$ (bottom) bin, assuming the G1a-4.5 and KOF formulas. While most clouds are low-mass, half the total molecular mass is contained in 490
clouds with $M_{\rm mol} > 6.8 \ee5$ \msun\ 
and half the total star formation occurs in 
192 clouds with $M_{\rm mol} > 1.3\ee6$ \msun, 
as more massive clouds tend to have lower $\alphavir$
(figure 2a of \citealt{2021ApJ...920..126E}).
The most probable virial parameter is 
$\sim 4$, similar to the value found for nearby galaxies in the PHANGS-ALMA survey \citep{2020ApJ...901L...8S}.
The fraction of molecular mass contained in bound clouds ($\alphavir < 2$) amounts to only 
17\% 
(see also \citealt{2021ApJ...920..126E}), but they account for 
54\%
of the total star formation.
Our model prediction that most star formation occurs in the most massive clouds (with relatively low $\alphavir$) is consistent with the observational finding that half 
the star formation 
in the Galaxy occurs in a small number of the most massive star-forming complexes
\citep[e.g.,][]{2010ApJ...709..424M}.

\subsection{Caveats and Further Work}\label{caveats}

Our success in predicting the observed SFR of the Galaxy rests on 4 pillars:
the MD catalog of molecular structures, the dependence of metallicity on \rgal,
the variation of \alphaco\ with $Z$, and the dependence of \epsff\ on
\alphavir.

The MD catalog is the only one to account for all the CO emission in the
Galaxy, so it is the only one suitable for prediction of the SFR of the
Galaxy. Only a small fraction of the mass in that catalog is in bound
structures ($\alphavir \leq 2$), with fractions varying from 0.07 for
Gong1a to 0.39 for LD. However, other methods of structure identification
\citep[e.g.,][]{2016ApJ...822...52R}, 
while also finding that most structures are unbound, 
find a much higher fraction of the molecular mass in bound structures.
Other structure identification procedures should be attempted and compared
to the observations of SFR.
 
Other transitions and isotopologues can also help to separate
the effects of luminosity-to-mass conversions from those of efficiency.
Structures defined by the \jj10\ or \jj21\ transitions of \coo\ were closest to the boundary between bound and unbound
\citep{2021ApJ...920..126E},
and full Galaxy surveys of these could provide new tests of theory if the
appropriate simulations of their emission are available.

Alternatively, tracers that clearly favor bound
structures, such as millimeter-wave continuum, or HCN emission, could be used
to survey the entire Galaxy. 
These tracers predict star formation rates with lower dispersion than
does CO
\citep{2016ApJ...831...73V,2019ApJ...880..127J}
and the resulting $\epsff \approx 0.01$ with a dispersion between studies of
$ \approx 0.3$ dex
\citep[and references therein]{2019ARA&A..57..227K}.
These can separate the effects of varying \alphaco\ from those of \epsff,
but require determinations of the HCN to dense gas mass conversion factor
($\alpha_{\rm HCN}$). This has been calculated from simulations by
\citet{2018MNRAS.479.1702O}, who find $\alpha_{\rm HCN} = 14\pm 6$ \alphacounit,
in reasonable agreement with observational constraints
\citep{2005ApJ...635L.173W, 2017A&A...604A..74S, 2020ApJ...894..103E}. 
However, variations in $\alpha_{\rm HCN}$ with environment
\citep{2017A&A...604A..74S} and $Z$ should also be considered based
on comparison to outer Galaxy clouds (S. Patra, personal communication),
and the density ``traced" by the \jj10\ transition of HCN may be substantially
lower than usually thought when the total HCN luminosity is considered
\citep{2020ApJ...894..103E}. 

The radial dependence of the SFR should be improved using modern surveys, such as
Hi-GAL
\citep{2010PASP..122..314M}
for infrared-based SFR. Results from two lines of sight look promising
\citep{2017A&A...599A...7V}.
An alternative method, with the advantage of velocity information from
recombination lines, would use more complete surveys of \hii\ regions
\citep{Anderson:2011, Anderson:2012, Anderson:2014}.
Both the inner and outer Galaxy provide strong tests of the dependence of
\alphaco\ on \rgal; the outer Galaxy has the advantage of less confusion
and deserves more attention.

The dependence of $Z$ on \rgal\ that we assumed 
is not well constrained inside about 5
kpc, where much of the molecular emission arises. Determination of gradients
at smaller \rgal\ would be extremely important. There are also hints at 
azimuthal variations in $Z$
\citep{2019ApJ...887..114W},
which might be incorporated as knowledge of distances continues to improve.
The gradient we assumed is the smallest of several choices, that range up
to $-0.077$ dex/kpc for C/H. These steeper Z gradients produce shallower
variations of $\Sigma_{\rm SFR}$ with \rgal, and these do not match as
well the current observations. Once improved observations of both abundances
and $\Sigma_{\rm SFR}(\rgal)$ are available, stronger tests will be possible.

The simulations of 
\citet{2020ApJ...903..142G}
covered $Z = 0.5$--$2.0$. As can be seen from Figure~\ref{acomodels},
we are mildly extrapolating those results in the very outer Galaxy and,
more importantly, in the inner Galaxy. Extending the simulations over
a wider range of $Z$ would be very valuable 
(see, e.g., \citealt{2022arXiv220103885H}).

The simulations that resulted in our assumed dependence of \epsff\ on
\alphavir\ can also be extended. In addition, the issue of how to assign
an observed \alphavir\ to a simulation adds some uncertainty.
These calculations assume that the current properties represent those at the
start of the simulation, that the fit to the simulations can be used for any
\alphavir\ in the observations, and, of course, that the simulations reflect
reality. While the clouds in the \citet{MD17}
sample are in a wide range of evolutionary states, the simulations suggest that the current observed values of \alphavir\ will not overestimate the initial values substantially; in fact \alphavir\ decreases from the initial value over the first  4 Myr (figure 5 of
\citealt{2021ApJ...911..128K}). Those simulations cover a range of 
\alphavirz\ of 1 to 5, and we have added a model with $\alphavirz = 10$.
While the observed \alphavir\ ranges from 0.1 to 100, there are few clouds with $\alphavir < 1$ but about one-third of the mass is in clouds with $\alphavir > 5$. Based on the new simulation with $\alphavirz = 10$, the expected star formation
contribution of such clouds is very low (Figure \ref{histplot}).

We reiterate that the method used 
to identify and characterize molecular structures is important. In particular, the definition of the size of the structure and assignment of mass to a certain size region strongly affects the density and virial parameter. Simulations of observables from the theoretical simulations will
help to identify the best method to assign sizes for comparison to the
theoretical models.

\section{Conclusions}

Accounting for a metallicity-dependent factor to convert CO luminosity to mass
and a virial-parameter-dependent star formation efficiency can bring theoretical predictions of the star formation rate into alignment with observed values 
for the Milky Way. 
While both play a role, the virial-parameter dependence of the star formation efficiency has the larger effect.
We also predict the variation of the star
formation rate with Galactocentric radius for different models of
$\alphaco(\rgal)$, which can be compared to improved determinations of the
observed variation. 
These will be most strongly affected by the metallicity-dependence of
the conversion from CO luminosity to mass.

\begin{acknowledgments}
We thank the referee for suggestions that improved the clarity of 
the paper.
We thank H. Dinerstein, K. Arellano-C{\'o}rdova, and C. Sneden 
for informative discussions on abundance 
measurements.
NJE thanks
the Department of Astronomy at the University of Texas
at Austin for ongoing research support.
J.-G.K. acknowledges support from the Lyman Spitzer, Jr. Postdoctoral Fellowship at Princeton University.
J.-G.K. acknowledges financial support from the EACOA Fellowship awarded by the East Asia Core Observatories Association. The work of ECO was partly supported by the
National Science Foundation (AARG award AST-1713949).
\end{acknowledgments}

\software{{\tt astropy} \citep{2013A&A...558A..33A,2018AJ....156..123A}, {\tt IPython} \citep{Perez07}, {\tt matplotlib} \citep{Hunter07}, {\tt NumPy} \citep{vanderWalt11}}

\bibliographystyle{aasjournal}

\bibliography{cite}

\end{document}